\documentclass[letterpaper,twocolumn,10pt]{article}
\usepackage{usenix-2020-09}

\usepackage{booktabs,xcolor,hyperref,enumerate}
\usepackage[T1]{fontenc}
\usepackage[utf8]{inputenc}
\usepackage{bm,courier,xspace,fancyhdr,fancyref,wrapfig}
\usepackage{amsmath,amsfonts,tabularx,tablefootnote,wrapfig,hyphenat}
\usepackage{mathtools}
\usepackage[labelformat=simple,skip=0pt]{subcaption}
\usepackage{xfrac,algorithm,algpseudocode}
\usepackage{paralist,mathtools,amsmath,upgreek}
\usepackage{float,enumitem,lipsum}
\usepackage[capitalize]{cleveref}
\usepackage{MnSymbol}
\usepackage{wasysym}
\usepackage[english]{babel}
\usepackage{blindtext}
\usepackage{times}

\newcommand{\textred}[1]{\textcolor{red}{#1}}
\ifx\noeditingmarks\undefined
   \newcommand{\pgwrapper}[2]{\textred{#1: #2}}
\else
   \newcommand{\pgwrapper}[2]{}
\fi

\DeclareCaptionLabelSeparator{emdash}{--- }

\newcommand{\systemname}{MMGaP}
\newcommand{\systemnames}{MMGaP's}
\newcommand{\sysname}{MMGaP}
\newcommand{\sysnames}{MMGaP's}

\title{MMGaP: Multi-User MIMO Detection and Precoding using GPU-assisted Physics-inspired Computation}

 \author{Abhishek~Kumar~Singh, Kyle Jamieson \\ Department of Computer Science, Princeton University}

\begin{document}
\maketitle

\begin{abstract}
Physics-inspired and quantum compute based methods for processing in the physical layer of next-generation cellular radio access networks have demonstrated theoretical advances in spectral efficiency in recent years, but have stopped short of practical realization on commodity processors, leaving a gap between the throughput practical systems can achieve and the projected throughput the state-of-the-art should achieve. To fill this gap, this paper proposes \textbf{MMGaP}, an uplink multi-user MIMO detector and downlink Vector perturbation precoder for next-generation cellular networks. MMGaP realizes these large MIMO processing algorithms for the first time on bare-metal CUDA kernels that scale to run on large GPU processing platforms, and can be packaged as TensorFlow modules, allowing easy integration with a variety of systems.  We integrate MMGaP with NVIDIA’s software-defined, GPU-accelerated 5G platform and evaluate its performance against the state-of-the-art. In a 5G cellular network using 100 MHz of radio bandwidth, eight antennas at the base station and eight concurrent users, we show that MMGaP improves uplink throughput by approximately 50 Mbps per user and downlink throughput by 100 Mbps per user over a wide range of SNR. We further show that MMGaP can also support larger MIMO sizes: for 16 antennas at the base station and 16 concurrent users, MMGaP provides more than 50 Mbps higher uplink throughput per user. We measure the execution time of MMGaP on different NVIDIA GPUs and show that it can operate at line-rate and meet the timing requirements of state-of-the-art 5G systems. 
\end{abstract}

\section{Introduction}
The internet has been historically dominated by downlink traffic---the ever-increasing popularity of video streaming platforms maintains a high demand for downlink data even today. The last decade has also seen a tremendous increase in interactive video for work and school, with platforms like Zoom, Google Meet, and Microsoft Teams becoming an important part of our day-to-day lives. Further, online gaming, virtual and augmented reality, and the use of other interactive web services continue. This has increased demand for high data rates in the uplink as well, and state-of-the-art cellular networks need to support high data-rates in both directions. 
\begin{figure*}
\centering
\includegraphics[width = \linewidth]{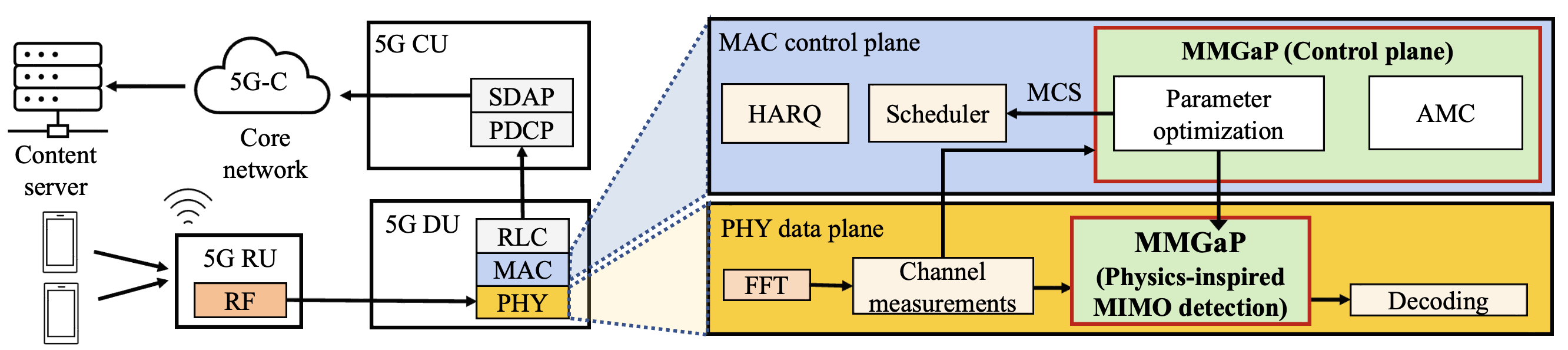}
\caption{High Level Design (Uplink): Our proposed system \systemname~is part of the Physical layer data plane (to perform MIMO detection) and MAC Layer control plane (for selecting best parameters, modulation, and coding schemes) in the 5G system.}
\label{fig:design}
\end{figure*}

While the demand for high-speed cellular internet is increasing, practical 5G deployments and other popular real-life systems like  Agora~\cite{agora}, Argos~\cite{argos}, BigStation~\cite{bigstation}, Hydra~\cite{hydra}, and LuMaMi~\cite{lumami} still use old and simple MIMO detectors/precoders like Minimum mean-square error~(MMSE) and Zero Forcing~(ZF)~\cite{mimoReview}. Instead of improving the MIMO detection/precoding algorithms, these practical systems employ linear detection and precoding methods that only perform well when the ratio of the number of transmit antennas and the number of receive antennas is very high, \textit{i.e.}, \emph{massive MIMO}, characterized by well-conditioned channels that allow linear methods to achieve good performance~\cite{mmseMassiveMimo1,mmseMassiveMimo3}. However, the trade-off is under-utilization of the system, as ideally the base station can simultaneously support as many users as the number of antennas at the base station. Furthermore, it is well known that the capacity of a MIMO system is bounded by the minimum among the number of transmit/receive antennas~\cite{Tse05fundamentalsof}; as a consequence, allowing more users to communicate simultaneously with the base station can improve the overall cell capacity.  However, to transition from Massive MIMO to Large MIMO (number of transmit antennas are equal to the number of receive antennas), base stations need to deploy more sophisticated MIMO detectors/precoders, as the performance of linear methods, such as MMSE, deteriorates significantly when the number of transmit antennas is equal to the number of receive antennas~\cite{performanceMMSE}. The optimal MIMO detector is known to be the Maximum Likelihood MIMO~(ML-MIMO) detector, however, it requires solving an NP-Hard optimization problem and its exponential complexity makes it infeasible for real systems with many users and antennas~\cite{ri-mimo}. Similarly, NP-Hard precoding methods like Vector Perturbation Precoding~(VPP) are known to provide much higher throughput than linear methods~\cite{vpPaper}. Therefore, enabling ML-MIMO detection/VPP for practical systems is a key step towards improving the overall throughput. 

Recent years have seen significant strides in the use of ``Physics-inspired'' computation for solving tough computational problems. Physics-inspired methodologies include Quantum Annealing~\cite{hauke2020perspectives}, digital-circuit Ising solvers~\cite{yamaoka201520k,aramon2019physics,goto2019combinatorial,goto2021high,leleu2020chaotic}, Oscillator-based Ising machines~\cite{oim}, Coherent Ising machines~\cite{wang2013coherent,marandi2014network,mcmahon2016fully,inagaki2016coherent,dopo,oeo}, spintronic and memristor Ising machines~\cite{sutton2017intrinsic,grollier2020neuromorphic,cai2020power}, bifurcation machines~\cite{tatsumura2021scaling}, and photonic Ising machines~\cite{roques2020heuristic,prabhu2020accelerating,babaeian2019single,pierangeli2019large}. Such methods have emerged as a powerful alternative to conventional algorithms, and significant theoretical performance gains have been demonstrated for several tough computational problems like MAXCUT~\cite{isingMaxCut}, Graph Coloring~\cite{isingGraphColor}, SAT~\cite{dwaveSat}, channel coding~\cite{srikar_ldpc,srikar_polar,qa_rxo_ldpc}, ray-tracing~\cite{qa_rayTracing}, and our problems of interest: MIMO detection~\cite{di-mimo, mdimimo,ri-mimo,minsung2019,minsungParallelTemp,kim2022warm,pSuccessQA} and Precoding~\cite{qavp}. However, these works are either based on computer simulations or utilize real implementations that are severely limited by their form factor and programming latency (time taken to set up a problem on the device). Therefore, despite the theoretical gains demonstrated by these methods, their feasibility for a practical 5G system remains unclear. Although these methods continue to grow and overcome their shortcomings rapidly, they are still at least a decade away from surpassing silicon-based systems~\cite{kasi2021challenge}.

An attractive and quicker way to extract the benefits of "Physics-inspired" computation is to instead create a silicon-based solver that emulates the dynamics of coherent Ising machines (CIM). 
Delta-Ising MIMO has shown theoretical performance gains for MIMO systems with a large number of users and high-order modulations, and outperforms other physics-inspired algorithms for MIMO detection \cite{di-mimo,mdimimo}, but is a Matlab simulation-based study and does not account for the timing and processing constraints of the 5G Radio Access Network (RAN). Thus, developing a 5G RAN-compliant CIM-based MIMO detector/precoder that meets the timing and processing deadlines remains open.

In this direction, this paper presents \textbf{\systemname{}}, a GPU based, Physics-inspired MIMO detector/precoder for uplink/downlink MU-MIMO, and is the first efficient realization of a physics-inspired MIMO detection/precoding that conforms to 5G standards and meets the processing and latency requirements of a practical 5G system. We have implemented \systemname{} on custom bare-metal CUDA kernels packaged as a TensorFlow library that can utilize the parallel processing capabilities of GPUs. 

To demonstrate the feasibility and performance gains of \systemname~for real-world 5G networks, we integrate \sysname{} with NVIDIA's software-defined, GPU-accelerated 5G framework called \textit{Aerial CUDA}.  
Our performance evaluation measures the improvement in physical layer throughput \systemname~provides over linear methods like the MMSE detector and Zero-forcing~(ZF) precoder. Note that linear MIMO detectors, such as MMSE and ZF, are not only popularly used by industry but also popular MIMO testbeds like Agora~\cite{agora}, Big-station~\cite{bigstation}, Hydra~\cite{hydra}, and LuMaMi~\cite{lumami}  for uplink data transmissions. Since it is well known that MMSE performs better than ZF~\cite{mmse_zf_1} we compare \systemname{} against MMSE in the uplink. Additionally, we also compare \systemname{} agaisnt the popular MMSE-SIC~\cite{optimalMMSEsic} MIMO detector. We show that, for a $8 \times 8$ and $16 \times 16$ 5G MU-MIMO system with 100 MHz bandwidth and 30 KHz subcarrier spacing,  \systemname{} can improve the physical layer throughput by approximately 50 Mbps per UE over a wide range of signal-to-noise ratios (SNRs), raging from 10~dB to 40~dB. In the downlink, for a $8 \times 8$ MIMO system, \systemname{} can improve the physical layer throughput by approximately 100 Mbps per UE over a wide range of signal-to-noise ratios. We also provide extensive microbenchmarks on the execution time of \systemname{} on various NVIDIA GPUs and demonstrate that it can meet the processing requirements of state-of-the-art 5G systems. 

This paper addresses a critical limitation of 5G systems: dependency on linear detection/precoding methods leading to highly suboptimal throughput performance. In this paper, we harness physics-inspired computation to enable nonlinear detection/precoding for 5G systems (leading to significant throughput improvements), and provide GPU-based design and implementation that can meet the timing requirements of real-world 5G deployments.
\begin{figure*}
\centering
\includegraphics[width = \linewidth]{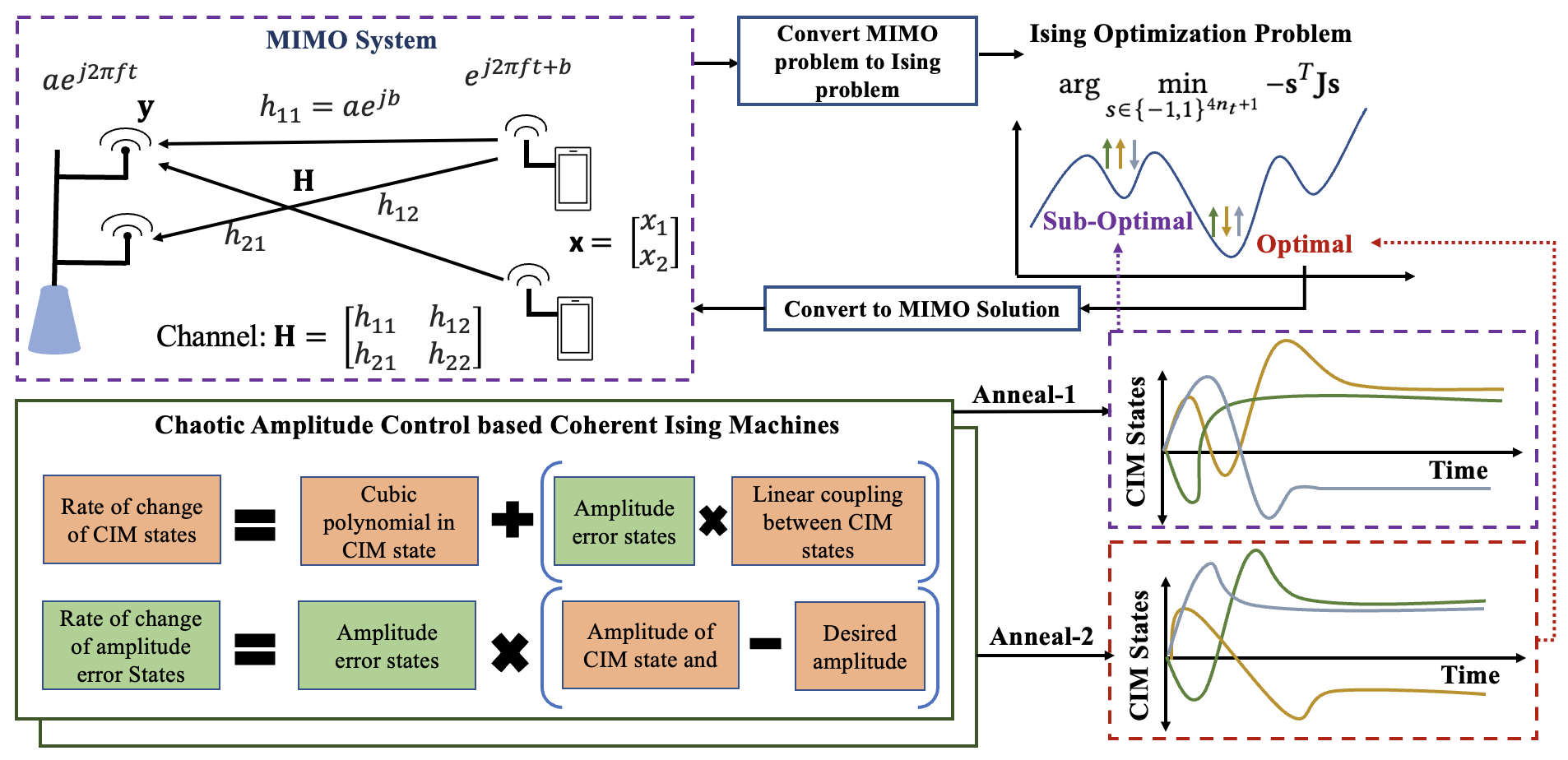}
\caption{Coherent Ising Machine-based MIMO detection: The MIMO detection problem for the Uplink MU-MIMO system is converted into an Ising optimization problem. The Ising problem is solved using the CIM-CAC algorithm (simplified working equations in the bottom left). We solve the same problem instance multiple times called ``anneals". In the bottom right, we see that each anneal leads to a different convergence state (due to randomness in the initialization of CIM-CAC). These solutions can correspond to sub-optimal solutions to the Ising problem or the optimal solution (top right).  We select the best solution found by CIM-CAC and convert it back to the corresponding MIMO solution.}
\label{fig:primer}
\end{figure*}
The rest of the paper is organized as follows: \S\ref{sec:primer} provides a background on MIMO system model and Coherent Ising machines. \S\ref{s:design} presents the design of \systemname{}, and \S\ref{s:impl} describes its implementation. We present an extensive performance evaluation of our proposed method in \S\ref{s:eval} for a state-of-the-art 5G scenario.

\section{Primer}
\label{sec:primer}
In this section, we describe the multi-user MIMO system and provide a brief background on Coherent Ising machines.
\subsection{Uplink Multi-user MIMO~(MU-MIMO)}
We consider an uplink multi-user MIMO~(MU-MIMO) system with $n_r$ antennas at the base station and $n_t$ users (with a single antenna). The channel is represented by an $n_r \times n_t$ channel matrix $\mathbf{H}$. The accumulation of all transmit symbols is represented by the transmit vector $\mathbf{x}$, where each element of $\mathbf{x}$ is drawn from a fixed constellation $\Omega$. The received signal is given by 
\begin{equation}
    \mathbf{y} = \mathbf{H}\mathbf{x} + \mathbf{noise},
\end{equation}
where $\mathbf{noise}$ is white Gaussian noise. The maximum-likelihood MIMO detector is given by,
\begin{equation}
    \mathbf{x}_{ML} = \arg \min_{\mathbf{x} \in \Omega^{n_t}}||\mathbf{y} - \mathbf{H}\mathbf{x}||^2
    \label{eq:ml}
\end{equation}

While the maximum likelihood detector is theoretically optimal, it has exponential complexity and is intractable for systems with many users. As a result, practical systems resort to using MMSE detector given by 
\begin{equation}
    \mathbf{x}_{MMSE} = (\mathbf{H}^\dag\mathbf{H} + \rho\mathbf{I})^{-1}\mathbf{H}^\dag \mathbf{y},
\end{equation}
where $\rho = \frac{E[|\mathbf{noise}|^2]}{E[|\mathbf{x}|^2]}$.

\subsection{Downlink Multi-user MIMO}
Similar to uplink,  we consider system with $n_t$ antennas at the base station and $n_r$ users (with a single antenna). The channel is represented by an $n_r \times n_t$ channel matrix $\mathbf{H}$. The accumulation of all transmit symbols is represented by the transmit vector $\mathbf{x}$. The received signal is given by 
\begin{equation}
    \mathbf{y} = \mathbf{H}\mathbf{x} + \mathbf{noise},
\end{equation}

However, unlike the uplink, a user is only aware of its own channel and not the entire channel matrix $\mathbf{H}$ and they cannot decode the received signals. Therefore, in the downlink, the base station performs a precoding operation such that the transmit vector $\mathbf{x}$ is given by $\mathbf{x} = \mathbf{W}\mathbf{u}$, where $\mathbf{W}$ is the precoding matrix, and $\mathbf{u}$ is the accumulation of intended transmissions for each user (each element of $\mathbf{u}$ is drawn from a fixed constellation $\Omega$). The goal of the precoding operation is to perform a ``pre-inversion'' of the channel, such that the signal received by user i ($y_i$) depends only on $u_i$ and its own channel. Zero forcing precoding is popularly used by the industry with
\begin{equation}
    \mathbf{W} = \mathbf{H}^\dag(\mathbf{H}\mathbf{H}^\dag)^{-1}.
\end{equation}

Assuming that the total power transmitted by the base station is constrained at $P$. The transmit signal with Zero-forcing precoding is
\begin{equation}
    \mathbf{x} = P\dfrac{\mathbf{H}^\dag(\mathbf{H}\mathbf{H}^\dag)^{-1}\mathbf{u}}{||\mathbf{H}^\dag(\mathbf{H}\mathbf{H}^\dag)^{-1}\mathbf{u}||}. 
\end{equation}

Therefore, the effective SNR is given by
\begin{equation}
    \dfrac{P}{E[||\mathbf{noise}||^2] \times ||\mathbf{H}^\dag(\mathbf{H}\mathbf{H}^\dag)^{-1}\mathbf{u}||}.
\end{equation}

In order to maximize effective SNR, Vector-Perturbation Precoding~(VPP)~\cite{qavp} defines the intended transmission as $\mathbf{u} + \tau\mathbf{v}$, where $\tau$ is a constant and $\mathbf{v} \in \mathcal{Z}^{N_{t} \times 1}$ is a vector of Gaussian integers. The goal is to select the best $\mathbf{v}$ to maximize the SNR. Therefore, the VPP problem is given by 

\begin{equation}
    \mathbf{v}^\star = \arg \min_{\mathbf{v} \in \mathcal{Z}^{N_{u} \times 1}} ||\mathbf{W}(\mathbf{u} + \tau\mathbf{v})||.
\end{equation}

\subsection{Coherent Ising Machines} 
Coherent Ising Machines~(CIMs) are alternative computing methods designed to find the optimal solution (ground state) of an Ising optimization problem, which is a quadratic optimization problem where each variable can take values $\pm 1$. 


Note that all variants of CIMs are designed to solve the Ising optimization problem. Mathematically, an Ising problem can be described as:
\begin{equation}
    \arg \min_{s \in \{-1,1\}^{4n_t + 1}} -\mathbf{s}^T\mathbf{J}\mathbf{s},
\end{equation}
where the elements of matrix $\mathbf{J}$ are real-valued problem coefficients (called Ising coefficients), and the vector $\mathbf{s}$ represent 
the problem variables that can only take values $\pm 1$ (called \textit{spin variables}). CIM's goal is to find the ground state of the Ising problem.
\begin{figure}
\centering
\includegraphics[width = \linewidth]{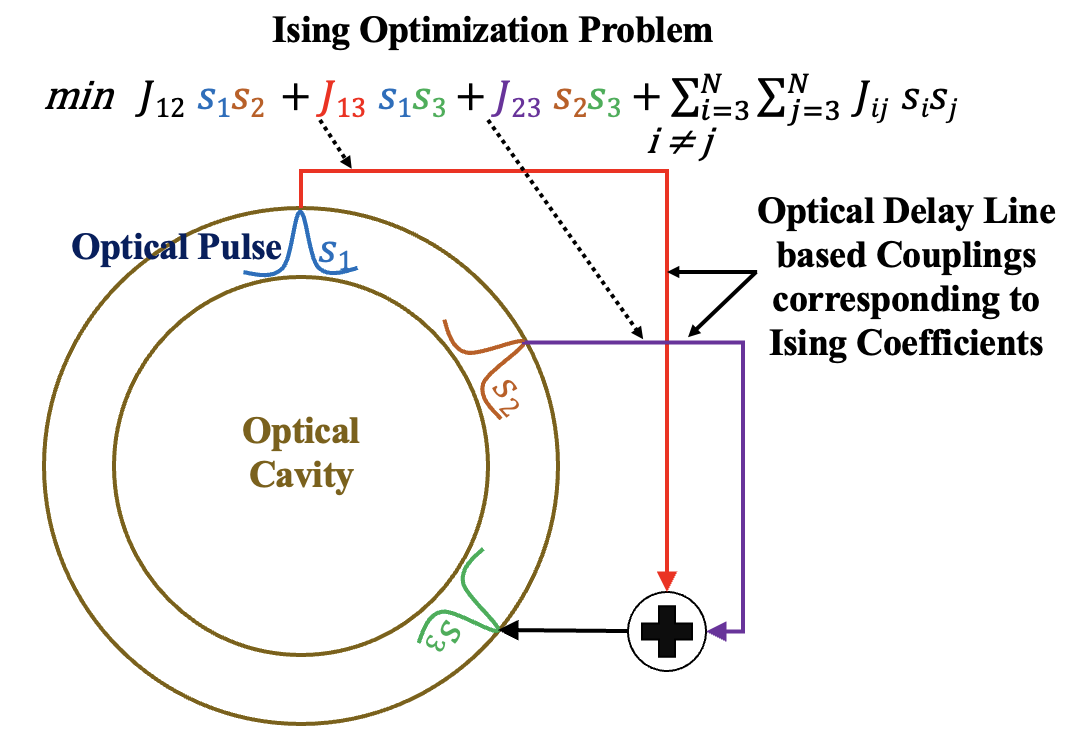}
\caption{Working principle of Degenerate Optical Parametric Oscillator based Coherent Ising Machine (DOPO-CIM): this example illustrates two couplings between pulses representing $s_3$ and those representing $s_1$ and $s_2$ (implementing the term $J_{13}s_{1}s_{3} + J_{23}s_{2}s_{3}$).}
\label{fig:cim_example}
\end{figure}
A DOPO-CIM~\cite{dopo} consists of optical pulses in a cavity and each pulse is used to implement one variable of the Ising optimization problem. The coefficients of the Ising problem are implemented via coupling between pulses implemented using optical delay lines (as illustrated in Figure~\ref{fig:cim_example}). Inspired by CIM's computational model, Chaotic-Amplitude control~(CAC) based CIM was suggested to improve the performance of the conventional CIM computational model~\cite{ampCorrectionCIM}. The intrinsic state of CIM-CAC is represented via real-valued \textit{state variables} $x_i$ where $i$ goes from 0 to number of variables in the problem minus one.
It introduces additional state variables, called \textit{error variables} ($e_i$), to regulate the amplitude of CIM's intrinsic state. It has been shown that CIM-CAC performs significantly better than the conventional CIM. The computational model of the Chaotic-amplitude control~\cite{ampCorrectionCIM} based Ising model~(CIM-CAC) represents each spin variable $s_i$ using a state variable $x_i$. The time evolution of these state variables is given by,
\begin{flalign}
 \forall i \text{      ,}\overbrace{\dfrac{dx_i}{dt}}^{\text{CIM State}} & = \overbrace{(p-1)x_i - x_i^3}^{\text{Cubic polynomial}} + \epsilon e_i \overbrace{\sum_{j\neq i} J_{i,j}x_j}^{\text{Coupling}}  \label{eq:cim1}
 \end{flalign}
 \begin{flalign}
  \forall i \text{,      } \overbrace{\dfrac{de_i}{dt}}^{\text{error states}} & = -\zeta(x_i^2 - a)e_i,\text{    }e_i > 0 \label{eq:cim2},
\end{flalign}
where $a,p$, and $\zeta$ are the free parameters of the model.

\textit{The usual procedure to solve the Ising optimization problem involves solving the same problem instance multiple times on the CIM and each run is referred to as an ``anneal''.} The rationale behind multiple runs is that the CIM can sometimes return a local minimum instead of the global optimal (as shown in Figure~\ref{fig:primer}); therefore, multiple anneals increase the probability of finding the ground state. For each anneal of CIM-CAC algorithm, in order to compute the solutions to the Ising problem, we need to numerically integrate Eq.~\ref{eq:cim1} and Eq.~\ref{eq:cim2} until convergence, and the final solution for each spin variable $s_i$ is given by $s_i = sign(x_i)$ at equilibrium. The number of anneals per problem instance ($N_a$) is a design parameter.




\section{Design}
\label{s:design}
\begin{figure*}
    \centering
    \includegraphics[width = \linewidth]{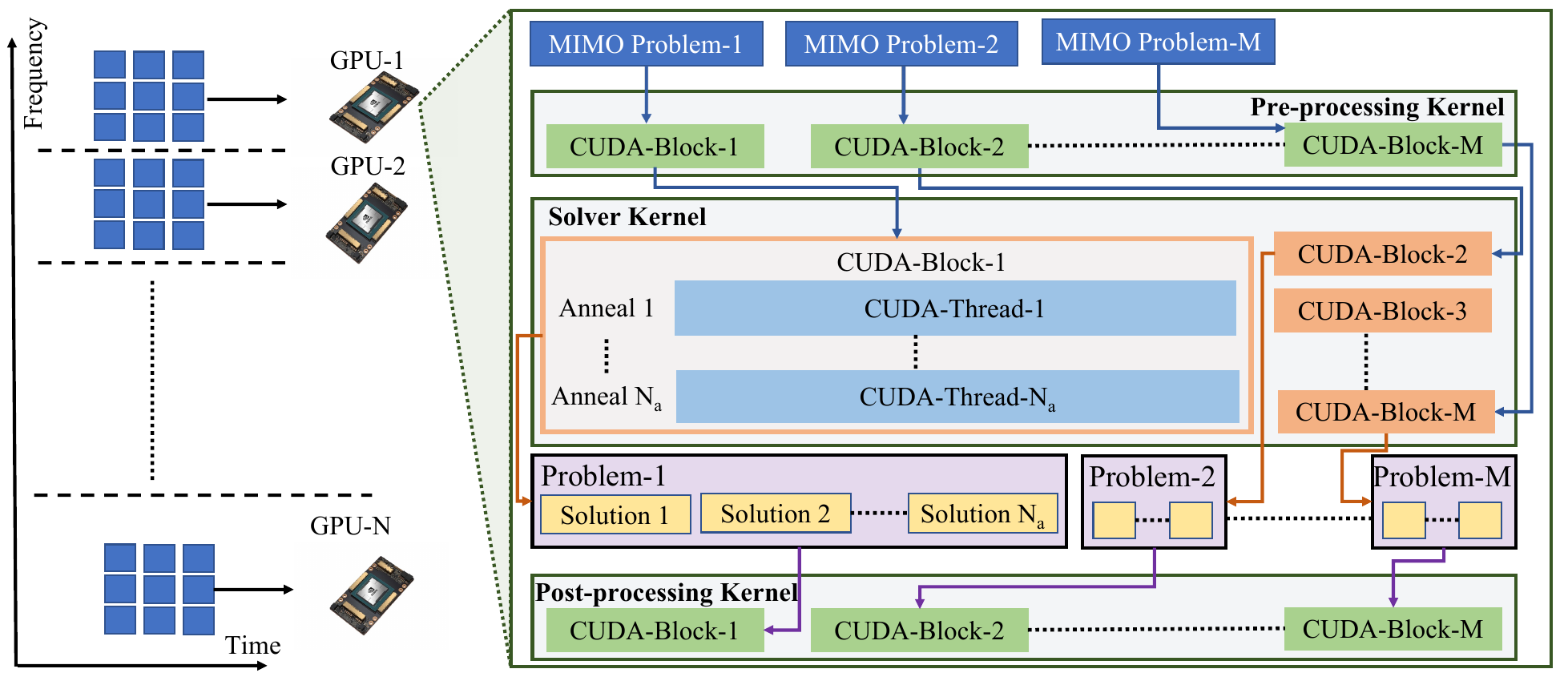}
    \caption{\textbf{\systemname{} processing flow:} the entire bandwidth is split into equal parts (equal to the number of available GPUs) and each part is assigned to a different GPU. Each MIMO problem is assigned to a different CUDA block within a GPU. Within a CUDA block, $N_a$ CUDA threads perform $N_a$ anneals (in parallel) corresponding to the assigned MIMO instance. }
    \label{fig:mmgap_gpu_design}
\end{figure*}
We begin with a description of
the structure of the 5G RAN (\S\ref{s:design:ran_integration}).
We next describe the processing flow of
\textbf{\systemname}, which is designed
using bare-metal CUDA kernels running on GPUs 
(\S\ref{s:design:processing_flow}). While we describe our design from the point of view of uplink MU-MIMO detection,  \S\ref{sec:mmgapdl} shows that the same design can be extended to downlink precoding with minimal changes. 

\subsection{RAN Integration}
\label{s:design:ran_integration}
A typical 5G RAN deployment is functionally split into three components: 
Radio Unit~(RU), Distributed Unit~(DU), and Centralized Unit~(CU). 
The RU is responsible for RF processing and is deployed close to the user. 
The DU hosts the Physical layer (PHY), MAC layer, and RLC layers. The 
PHY is responsible for recovering the transmitted bits from the 
baseband signals, MAC functionality includes controlling signaling
(including channel quality indicators~(CQI) and buffer status reports),
scheduling, grant allocation, and assigning the optimal modulation 
and turbo-code rate for each user's transmission. The RLC layer is 
responsible for ensuring the reliable transmission of data packets 
between the UE and the gNB. In an MU-MIMO 5G scenario, several 
users concurrently interact with a base station. Unlike 
conventional spatial multiplexing MIMO, where all MIMO layers are 
used to transmit data corresponding to a single user, MU-MIMO 
schedules each user on a separate MIMO layer. Since all users transmit
simultaneously, the signal received by the PHY of the base station 
is a linear combination of the data transmitted by each user:
MIMO detection separates these signals into respective MIMO layers,
then the base station processes each independently. 
This is followed by error-correction decoding, which if successful,
passes the data to higher layers. Otherwise, 5G uses a Hybrid 
Automatic Repeat Request~(HARQ) algorithm to retransmit. 

\begin{figure*}
    \centering
    \includegraphics[width = \linewidth]{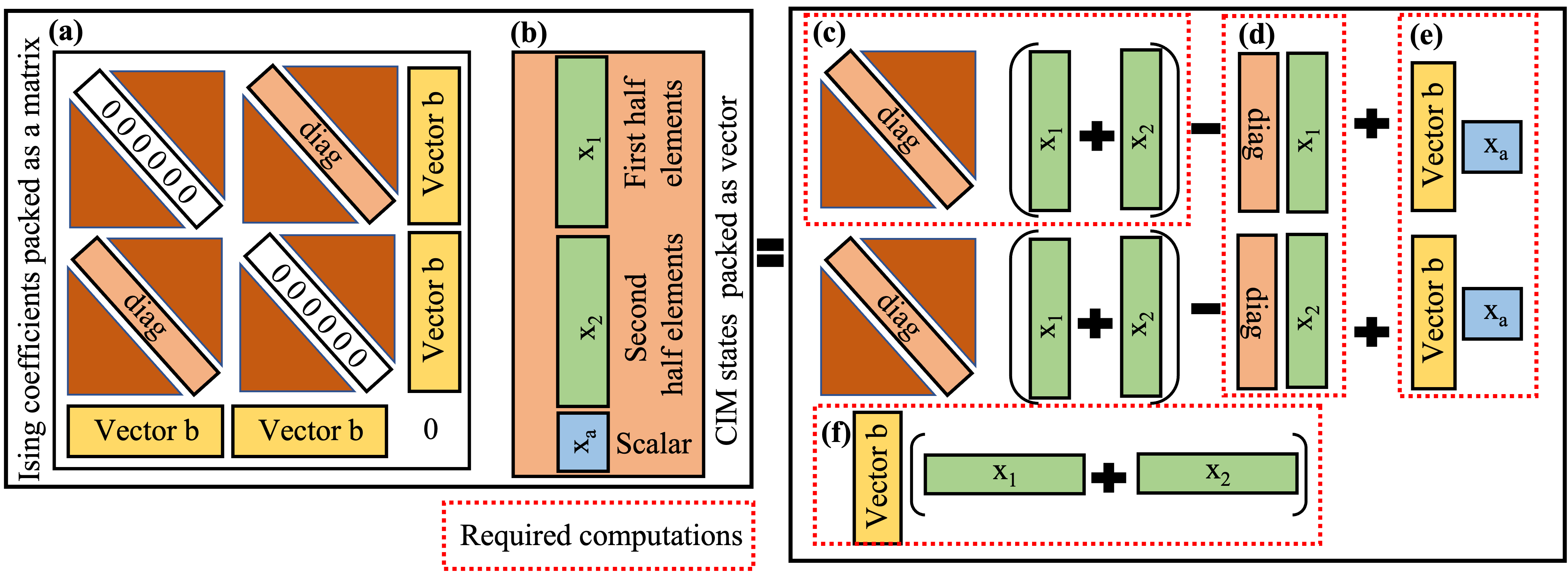}
    \caption{\textbf{\systemnames{} matrix-vector multiplication optimization:} Exploiting the internal structure of Ising coefficients generated from the MIMO detection problem allows a decomposition into a smaller matrix-vector multiplication and a few vector operations. (a) Ising coefficients stacked as a matrix, (b) CIM states stacked as two vectors and a scalar, (c) required MVM (of size half), (d) required element-wise products, (e) required scaling operations, and (f) required dot product}
    \label{fig:mvm_opt}
\end{figure*}
\subsection{Processing Flow}
\label{s:design:processing_flow}
The primary design goal of \sysnames{} processing flow is 
to efficiently divide the workload among the available computing resources,
to generate high-quality solutions 
within the timing and processing constraints of a 5G system. 
Figure~\ref{fig:mmgap_gpu_design} summarizes the high-level design:
a GPU running \systemname{} executes three CUDA kernels: the \textit{pre-processing kernel} which computes the Ising coefficients, 
the \textit{solver kernel} performs the
numerical integration of CIM-CAC for all sub-carriers in parallel, and the \textit{post-processing kernel}
selects the best solution found for each MIMO problem and 
generates the final solution. In addition to these, \systemname{} also includes functions to compute MMSE and MMSE-SIC solutions for the MIMO problems. These functions are implemented using a mix of bare-metal CUDA kernels and NVIDIA's CUBLAS library. All the functionalities provided by \systemname{} are also packaged as TensorFlow operations that can be accessed by
simple function calls.

\systemnames{} input
is a set of the following data elements, one for each MIMO problem:
\textbf{1)}~the wireless channel 
of the problem $\mathbf{H}$ and  \textbf{2)}~the received signal $\mathbf{y}$. This input data is processed using the MMSE function provided by \systemname{} to compute the MMSE solution ($\textbf{x}_{mmse}$).
As shown in Figure~\ref{fig:mmgap_gpu_design}, 
we split the entire frequency bandwidth of the 5G RAN into equal parts (numbering the number of available GPUs) with each part 
assigned to a different GPU.

With the input data ($\mathbf{H}$,$\mathbf{y}$) and the MMSE solutions (computed using the MMSE function) available, the first step is the \textbf{pre-processing kernel}, which is responsible for mapping the MIMO problem to an Ising optimization problem. Each resource element is assigned to a CUDA block that calculates the corresponding Ising coefficients ($J_{ij}$) corresponding to the MIMO problem associated with the resource element.  

In the \textbf{solver kernel}, the MIMO problem corresponding to a single resource element is assigned to a CUDA block. As the figure illustrates, each CUDA block consists of $N_a$ CUDA threads, each responsible for performing computations corresponding to a single anneal of the CIM-CAC algorithm. Since each anneal is inherently a sequential process due to the nature of numerical integration, distributing the processing of a single anneal among multiple GPU threads would be less efficient due to the need for constant synchronization. Therefore, a data-decomposition based approach, where each anneal is assigned to a different thread, is more appropriate and requires no synchronization between different threads.  The solver kernel considers the final state of the numerical integration process as a candidate solution. The solver kernel also calculates the ``energy'' of the final candidate solutions given by the ML objective function (\ref{eq:ml}).  Furthermore, if for an anneal the energy of the solution found is larger than the energy of the MMSE solution ($\mathbf{x}_{\mathrm{mmse}}$) then the MMSE solution is returned instead of the solution found by the anneal.


In the \textbf{post-processing kernel}, the solutions generated by the solver kernel are processed to find the best solution for each MIMO problem \textit{i.e.} the solution with the smallest energy. Each resource element is assigned to a CUDA block (with a single thread) which iterates over all the solutions generated for the MIMO problem and outputs the best solution.

\systemname{} detects the number of available GPUs and splits the
workload equally among them, dividing the entire 
frequency bandwidth into 
equal parts and assigning each to a different GPU. Once
all GPUs have finished processing the data, all the 
MIMO solutions are collected and sent for LDPC decoding. 
Note that since the bits corresponding to a single packet are 
spread across all allocated resource elements within the 
allocated grant, the LDPC decoding of the data
resulting from the MIMO problems in a given allocated
grant must be serialized after the detection of all such
problems.

On an NVIDIA GPU, a group of 32 threads, termed a \textit{warp},
is the fundamental unit of computation. All threads within a 
warp are executed at the same time (Single Instruction, 
Multiple Thread model). Therefore, since we are targeting
NVIDIA GPUs in this work, we choose $N_a = 32$, which allows
the processing of individual MIMO instances to be contained 
within a single warp and efficient parallel processing. 

\subsubsection{Matrix-Vector Multiplication Optimization}

The matrix-vector multiplication (MVM) required to compute the 
``coupling'' term in Eq.~\ref{eq:cim1} is the most computationally challenging task for \systemname{}. Since we utilize GPU parallelism to solve different MIMO problems in parallel, we cannot apply conventional GPU-accelerated MVM methods, as MVM needs to be performed by a single thread. \systemname{} utilizes the internal structure among the Ising coefficients generated from the MIMO problem to optimize the MVM calculation. 
\begin{figure}
    \centering
    \includegraphics[width = \linewidth]{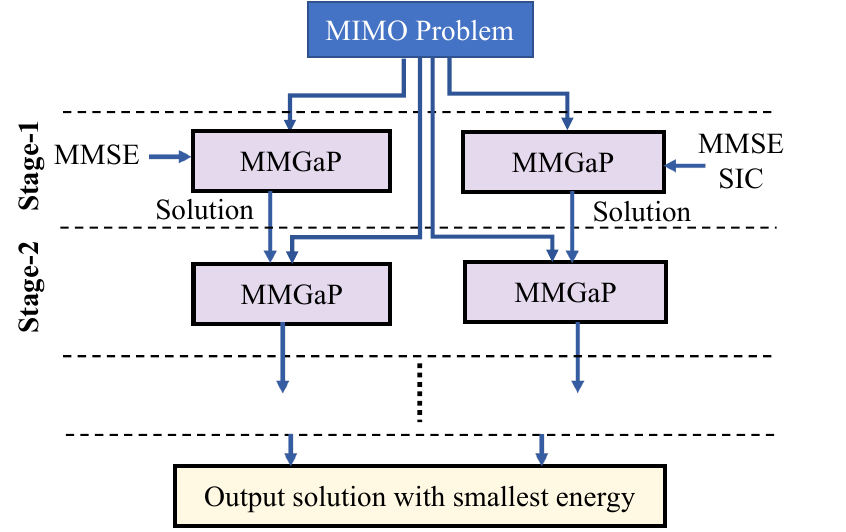}
    \caption{\systemname{}-E: Enhancing \systemname{} using a multi-stage, multi-seed approach. For each MIMO problem, two instances are created: one uses MMSE as the initial guess, whereas the other uses MMSE-SIC as the initial guess. Further, a multi-stage structure is used where the solution of the previous stage is used as the initial guess for the next stage.}
    \label{fig:multistage}
\end{figure}
As shown in Figure~\ref{fig:mvm_opt}, if we pack all the Ising coefficients corresponding to the MIMO problem as a matrix and look deeper into its structure, we see that the Ising matrix consists of a smaller matrix (about half the size) repeated twice on the diagonal (with its own diagonal zeroed) and repeated twice on the off-diagonal (see Figure~\ref{fig:mvm_opt}(a)). In addition, the last row and the last column each contain a vector \textbf{b} repeated four times (see again Figure~\ref{fig:mvm_opt}(a)). This structure is a consequence of the transform used to convert the MIMO problem into an Ising problem \cite{di-mimo}, which expresses both real and imaginary parts of a QAM symbol as a symmetric function of two spin variables. This leads to the symmetric nature of the Ising matrix. The diagonal of Ising matrix is zero because for any spin variable $s$, $s^2 = 1$ and therefore the diagonal terms (are constant) can be eliminated from the optimization process. 

 Instead of considering all state variables as one vector, we consider it as two vectors ($\mathbf{x}_1$ and $\mathbf{x}_2$ in Figure~\ref{fig:mvm_opt}(b)) stacked along with a scalar ($x_a$ in Figure~\ref{fig:mvm_opt}(b)). This allows us to express one large MVM as a combination of one MVM (Figure~\ref{fig:mvm_opt}(c)) of half the size, two element-wise products (Figure~\ref{fig:mvm_opt}(d)), two scaling operations Figure~\ref{fig:mvm_opt}(e), and one dot product Figure~\ref{fig:mvm_opt}(f). Note that the quantity $\mathbf{x}_1$ and $\mathbf{x}_2$ in Figure~\ref{fig:mvm_opt} also needs to be calculated only once and can be reused in other operations.

Note that the final solution of the CIM-CAC equation depends only on the sign of the CIM state variables. Therefore, the numerical integration process does not need to track the CIM-CAC equations exactly as long as the signs of the final states are the same. This allows us to further approximate Eq.~\ref{eq:cim1} and Eq.~\ref{eq:cim2}, by skipping the update to MVM at each iteration. We refer to the frequency at which MVM is updated as $f_{mvm}$. Another design parameter is the step size or the integration interval of the numerical integration procedure ($\Delta t$). Let us look at the trade-offs in the choice of $f_{mvm}$ and $\Delta t$: If $\Delta t$ is too small, it leads to slow growth of the differential equations and hence would require more iterations; on the other hand, if $\Delta t$ is too large, then the numerical integration process can be inaccurate and even fail to converge. If $f_{mvm} = 1$, \textit{i.e.} the MVM is updated at every iteration of numerical integration, then \systemname~would incur the high computation cost of MVM at each iteration (increasing its execution time). On the other hand, if $f_{mvm}$ is too large, then \systemname~will inaccurately track the CIM-CAC equations and would converge to the wrong final state. In fact, since MVM is the only term in the CIM-CAC equations that encapsulates the Ising coefficients, a large $f_{mvm}$ would imply that the optimization process ignores the objective function. For \textit{e.g.}, if $f_{mvm}$ is infinite, MVM is not calculated at all (except at the first iterations) and the state variables would evolve independently and would completely ignore the Ising coefficients. Based on empirical experimentation (Appendix~\ref{sec:num_int_micro} and Figure~\ref{fig:pdiv_heatmap},~\ref{fig:perr_heatmap}), we choose $f_{mvm} = 2$ and $\Delta t = 0.02$ for \systemname{}.

\subsubsection{\systemname{}-E: Performance Enhancement via Multi-stage, Multi-seed Approach} 
It has been shown that a multistage approach that uses multiple initial guesses instead of just one can significantly boost the performance of Ising machine-based MIMO detection~\cite{mdimimo}. In order to incorporate this into our system, for each MIMO problem, we create two independent Ising problem instances. As shown in Figure~\ref{fig:multistage}, one of them uses MMSE as the initial guess, while the other uses MMSE-SIC. The two instances are then solved using the \systemname{} solver. In the next stage, the solution generated for each instance is used as the initial guess for generating the Ising instance from the MIMO problem. This can be repeated for several stages, allowing \systemname{}-E to incrementally improve the quality of the initial guess used by the Ising machine-based MIMO detector. Furthermore, using multiple initial guesses allows our system to explore more of the optimization landscape. Finally, the overall best solution found is the final output. The processing chain from MMSE guess can be executed in parallel to the one starting from MMSE-SIC, however, in the current implementation (since \systemname{} aggressively uses the entire GPU) they are performed sequentially. It is only advantageous to run them in parallel if the workload is not enough to utilize the entire GPU (which will not be the case unless an extremely high number of GPUs are available). 
\begin{figure}
    \centering
    \includegraphics[width = \linewidth]{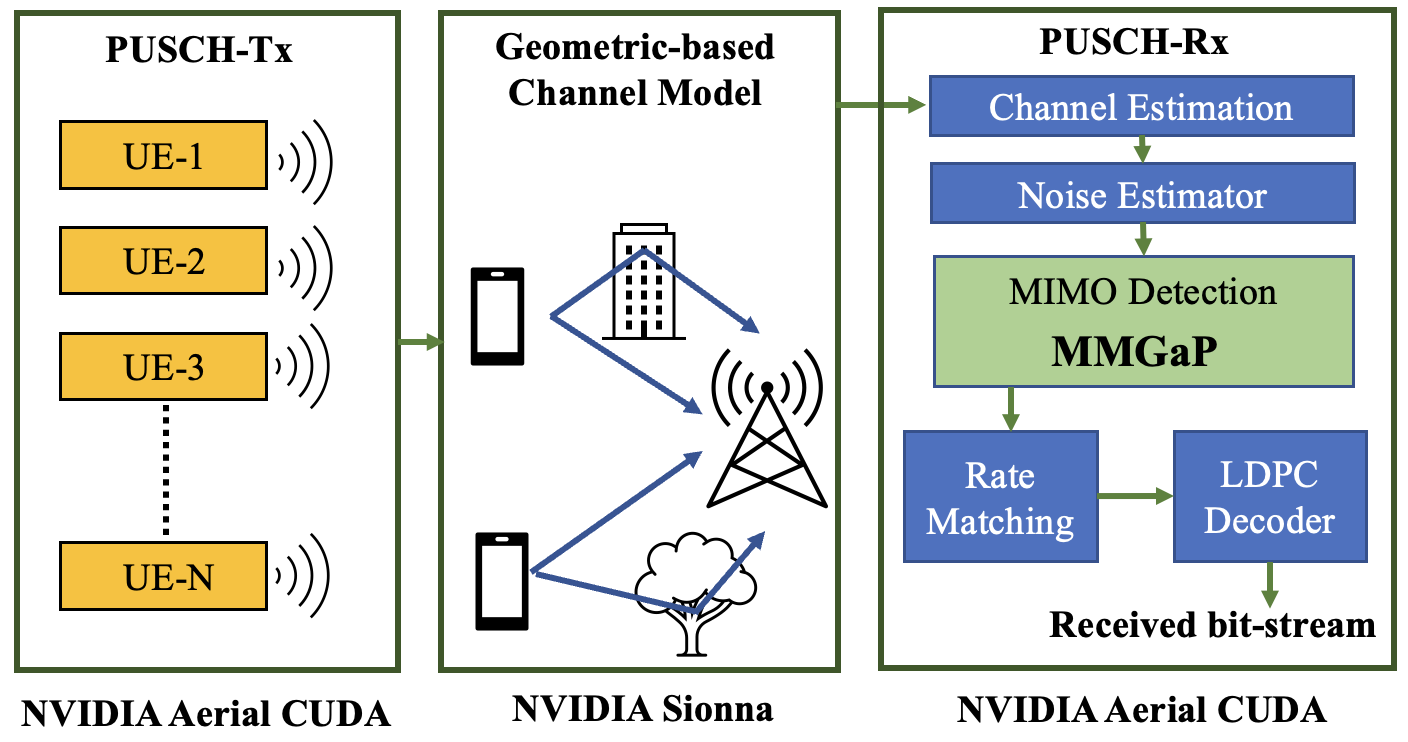}
    \caption{\systemname~Evaluation Methodology: The Uplink transmissions are based on NVIDIA's Aerial CUDA library. NVIDIA Sionna is used to generate a realistic, geometry-based channel model. The receiver processing is based on Aerial CUDA libraries implementing physical layer functionalities where the MIMO detector is replaced with \systemname.}
    \label{fig:eval_setup}
\end{figure}
\begin{figure}
    \centering
    \includegraphics[width = \linewidth]{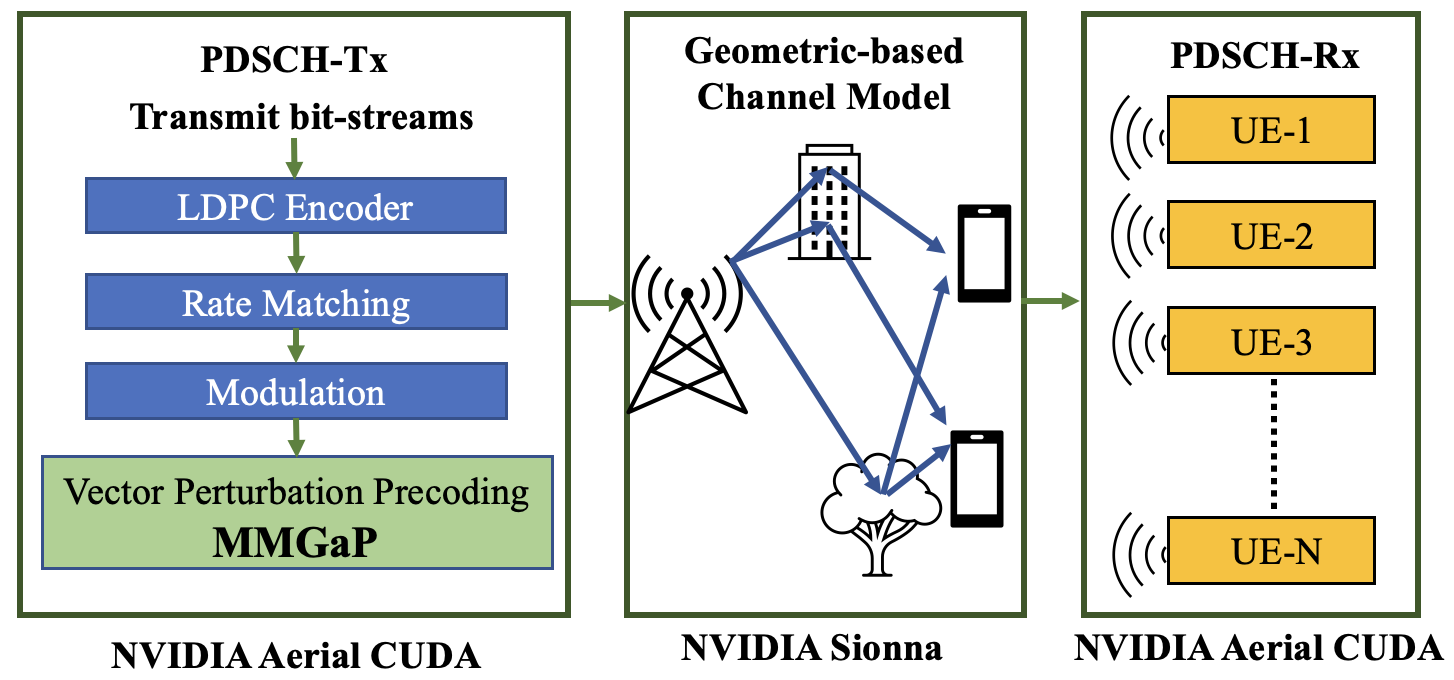}
    \caption{\systemname~Evaluation Methodology: The Downlink transmissions are based on NVIDIA's Aerial CUDA library. NVIDIA Sionna is used to generate a realistic, geometry-based channel model. The transmitter processing is based on Aerial CUDA libraries implementing physical layer functionalities where the Vector Perturbation Precoding is replaced added using \systemname{}.}
    \label{fig:mmgap_dl_method}
\end{figure}
\begin{figure}
    \centering
    \includegraphics[width=\linewidth]{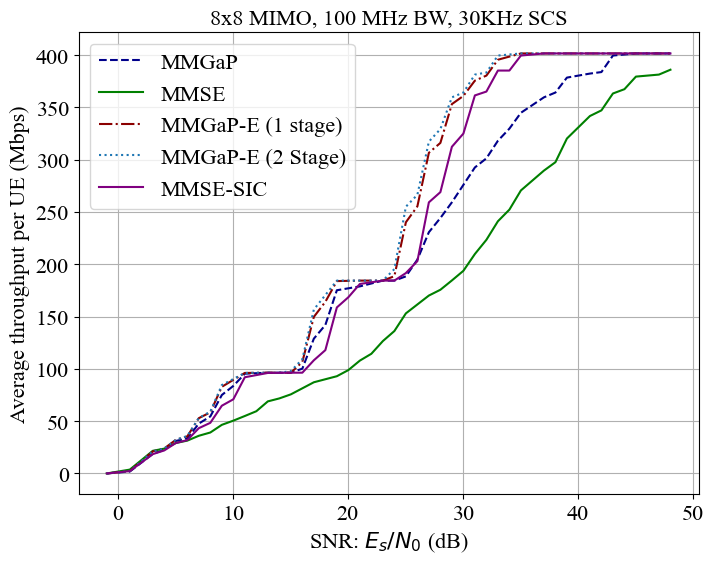}
    \caption{Uplink Physical layer throughput for $8 \times 8$ MIMO with 100 MHz bandwidth and 30 KHz SCS.}
    \label{fig:tput8x8_e}
\end{figure}
\subsection{Enabling  \systemname{} to Downlink}
\label{sec:mmgapdl}
Downlink MIMO is the exact opposite of the uplink scenario, where the base station will transmit to several UEs using multiple antennas. However, unlike uplink, where the base station receives signals from UEs and performs MIMO detection to decode their respective data streams, the UEs cannot perform MIMO detection because they do not know the entire channel matrix. While it is possible to communicate the entire channel to each UE, it will lead to significant overhead, and hence state-of-the-art wireless systems adopt a different approach. 

In the downlink, the base station performs a precoding operation to modify the transmitted signals so that the received signal at each UE $i$ is only a function of the channel corresponding to that UE (as described in \S\ref{sec:primer}). 

\systemname{} can also be used to enable Vector Perturbation Precoding~(VPP), an NP-Hard precoding methodology, to provide significant improvement in downlink performance. Next, we will demonstrate the  equivalence in the mathematical form of VPP and MIMO detection, which allows us to use the implementation of \systemname{} for downlink with trivial modifications.  

As described in \S\ref{sec:primer}, given the precoding matrix $\mathbf{W}$ and the data symbols $\mathbf{u}$, the VPP optimization problem is given by

\begin{equation}
    \mathbf{v}^\star = \arg \min_{\mathbf{v} \in \mathcal{Z}^{N_{u} \times 1}} ||\mathbf{W}(\mathbf{u} + \tau\mathbf{v})||.
\end{equation}

If we define $\tilde{\mathbf{y}} = \mathbf{Wu}$, and $\mathbf{H_p} = -\tau\mathbf{W}/2$, then the VPP problem is given by
\begin{equation}
    \mathbf{v}^\star = \arg \min_{\mathbf{v} \in 2 \cdot\mathcal{Z}^{N_{u} \times 1}} ||\tilde{\mathbf{y}} - \mathbf{H_p}\mathbf{v}||, 
\end{equation}
which is the exact same mathematical expression as the uplink MIMO detection problem and can be solved by \systemname{}. Therefore, our proposed system can also be used to enable Vector Perturbation precoding in the downlink. 

As noted before, \systemname{} requires the wireless channel of the problem $\mathbf{H}$, the received signal $\mathbf{y}$, and the MMSE solution $\mathbf{x}_{\mathrm{mmse}}$ for the uplink MU-MIMO problem. In case of a downlink precoding problem, given the mathematical equivalence, the input would be $\mathbf{H_p}$ (equivalent to the channel), $\tilde{\mathbf{y}}$ (equivalent to the received vector) and a vector of all zeros instead of the MMSE solution.

\section{Implementation}
\label{s:impl}
\begin{figure*}
    \begin{subfigure}{0.49\linewidth}
        \includegraphics[width=\linewidth]{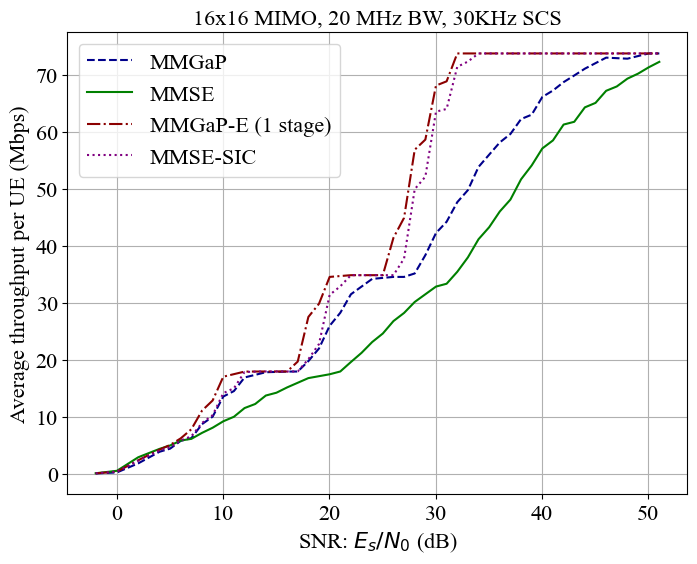}
    \end{subfigure}
    \begin{subfigure}{0.49\linewidth}
        \includegraphics[width=\linewidth]{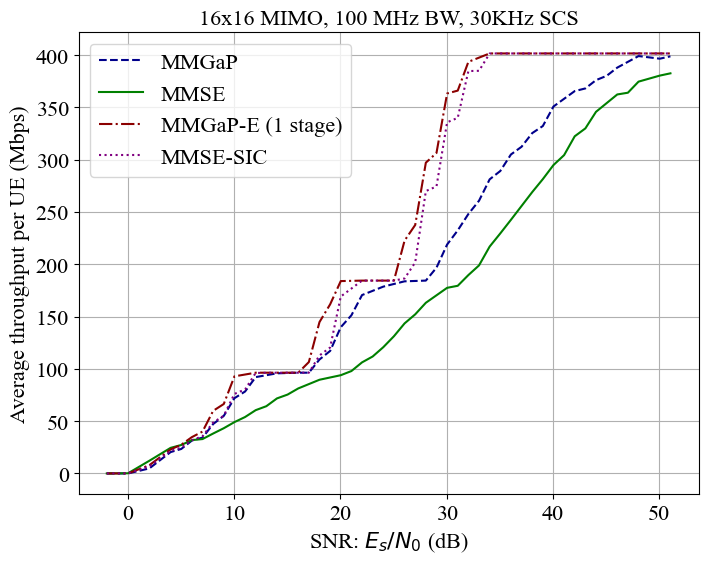}
    \end{subfigure}
    \caption{Physical layer throughput for $16 \times 16$ MIMO with 20 MHz and 100 MHz bandwidth and 30 KHz SCS.}
    \label{fig:tput16x16_e}
\end{figure*}


Our implementation consists of two parts: \systemname~solver packaged as Python modules, and its integration with NVIDIA's Aerial CUDA platform.
NVIDIA's Aerial CUDA is a virtualized software-defined, GPU-accelerated 5G stack. We integrate \systemname{} with the TensorFlow-based implementation of 5G Layer-1~(L1) present within Aerial CUDA (called PyAerial). PyAerial provides GPU-accelerated implementation of MIMO detection, LDPC encoding and decoding, rate-matching, and channel estimation. We implement an Uplink Multi-user MIMO scenario with PyAerial and modify the PUSCH processing chain to use \systemname{} instead of the conventional linear MIMO detection. We implement a Downlink Multi-user MIMO scenario with PyAerial and modify the PDSCH processing chain to use \systemname{} for downlink precoding. 

As shown in Figure~\ref{fig:eval_setup}, the Uplink data transmission or PUSCH-Tx for each UE is performed using libraries provided by Aerial CUDA. PUSCH transmissions are processed using a realistic geometry-based channel model generated using NVIDIA Sionna~\cite{sionna}. The received signals are processed using NVIDIA Aerial CUDA libraries where channel estimation, noise estimation, rate matching, and LDPC decoding are performed using Aerial CUDA libraries. However, we replace the default channel equalizer or MIMO detector with \systemname~allowing us to evaluate our method.

As shown in Figure~\ref{fig:mmgap_dl_method}, the Downlink data transmission or PDSCH-Tx for each UE is performed using libraries provided by Aerial CUDA. We introduce Vector Perturbation precoding using \systemname{} in the transmit flow. PDSCH transmissions are processed using a realistic geometry-based channel model generated using NVIDIA Sionna~\cite{sionna}. The received signals are processed using NVIDIA Aerial CUDA libraries.


We implement \systemname~using bare-metal CUDA. The CUDA kernels are packaged as a TensorFlow operation and can be accessed as a simple function call. This allows seamless integration with NVIDIA's platform. We integrate \systemname~with Python version of Aerial CUDA (PyAerial) and implement the Multi-user data path in TensorFlow (Figure~\ref{fig:eval_setup}). We run \systemname~on NVIDIA's A100 GPUs, however, \systemname~is compatible with any GPU that supports CUDA. 

\systemname{} also modifies the MAC control plane by changing the Adaptive Modulation and Coding~(AMC) feature. With 
\systemname{}, AMC is responsible for jointly optimizing the modulation, coding, and parameters of \systemname{} to maximize the link layer throughput. Currently, we use an Oracle AMC, and the design of a practical AMC will be addressed in the future.

\section{Evaluation}
\label{s:eval}
We evaluate \sysnames{} execution time and throughput in various experimental scenarios.

\subsection{Methodology}
We evaluate, \systemname~in an uplink multi-user MIMO 5G NR scenario: the base station has $N_r$ transmitter antennas, and there are $N_t$ mobiles (known as User equipment or UEs), each equipped with one transmit antenna, and positioned randomly in the cell. The UEs are assumed to be moving at 0.833 m/s (which would correspond to normal walking pace). The base station operates in Frequency Division FDuplexing (FDD) mode at 3.5 GHz. The uplink bandwidth (BW) is 100 Mhz with 30KHz sub-carrier spacing (SCS), which corresponds to 273 Physical resource blocks (PRBs) each containing 14 OFDM symbols and  12 subcarriers. Of the 14 OFDM symbols, 11 are used for transmitting data and one for transmitting the DMRS reference signal. No data is transmitted on the first two symbols of each RB. For downlink, all parameters remain the same, except that each UE is equipped with one receive antenna and the base station has $N_t$ transmit antennas.

Each UE is equipped with a single polarization antenna with an antenna pattern according to the 3GPP 38.901 specification. The base station is equipped with an dual cross-polarization antenna array with an antenna pattern according to the 3GPP 38.901. We assume an Urban Microcell (UMi) channel model in accordance with the 3GPP specification. Due to some inaccuracies with DMRS-based channel estimation (in multi-user MIMO scenario) employed by NVIDIA's Aerial CUDA platform, we assume an ideal channel estimation for this work.



\subsection{Physical Layer Throughput}
In this section, we look at the Physical layer throughput of \systemname{}, MMSE, and MMSE-SIC in the uplink; and \systemname{}-downlink and Zero-forcing precoding in the downlink. 
We employ a lookup table-based Adaptive Modulation and Coding (AMC) that contains the best MCS as a function of SNR for each algorithm.
\subsubsection{Uplink}
In Figure~\ref{fig:tput8x8_e}, we plot the physical layer throughput for an $8 \times 8$ MIMO system with 100 MHz bandwidth and 30 KHz subcarrier spacing. We see that \systemname~ provides significant performance improvements over the MMSE detector, and provides about 50 Mbps higher throughput over the majority of the SNR range. Only at very low SNRs (around 0 dB) the performance of MMSE and \systemname~ are similar. This can be attributed to the fact that even the theoretical performance of Ising machine-based MIMO detection is similar to MMSE at low SNRs~\cite{di-mimo,mdimimo}. We further see that \systemname{} can outperform MMSE-SIC low-mid SNR range, and falls behind for high SNR. This can be attributed to the fact that, at high SNR, base station  will se 64-QAM modulation and the limited search space around the single MMSE guess (used by \systemname{}) will only cover a small part of the constellation, and therefore is more likely to miss good quality solutions. As seen in Figure~\ref{fig:tput8x8_e}, \systemname{}-E solves and can provide provide significant performance improvements for MMSE-SIC as well. Further increasing the number of stages, from one to two, only slightly improves the performance of \systemname{}-E.


In Figure~\ref{fig:tput16x16_e}, we look at the throughput performance of our proposed methods for a $16 \times 16$ MIMO system. We see that \systemname{} can vastly improve the performance of the MMSE detector. Although MMSE-SIC performs better than \systemname{} for $16\times 16$ MIMO at high SNR, \systemname{}-E can provide significant improvements over MMSE-SIC.
\subsubsection{Downlink}
\begin{figure}
    \centering
    \includegraphics[width=\linewidth]{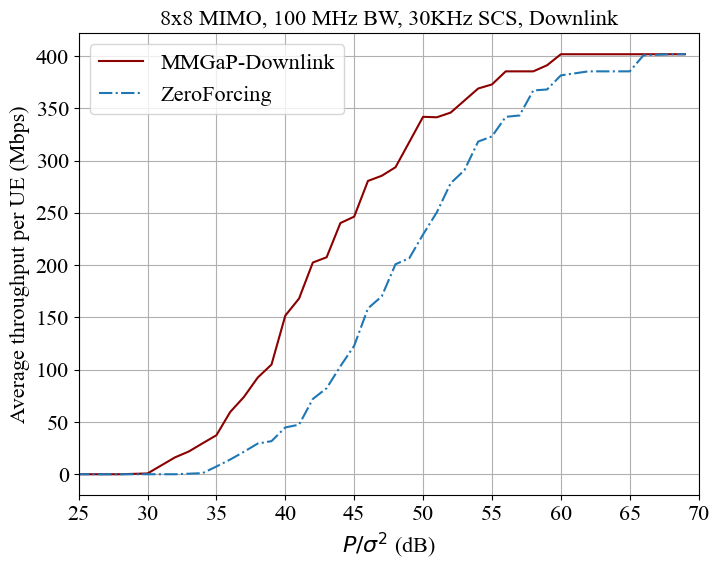}
    \caption{Downlink Physical layer throughput for $8 \times 8$ MIMO with 100 MHz bandwidth and 30 KHz SCS.}
    \label{fig:tput8x8_dl}
\end{figure}

In Figure~\ref{fig:tput8x8_dl}, we plot the downlink physical layer throughput for an $8 \times 8$ MIMO system with 100 MHz bandwidth and 30 KHz subcarrier spacing. We see that \systemname{} can provide significant performance improvements over the widely used Zero-Forcing precoding, and provide up to 100 Mbps higher throughput per user.

\subsubsection{High connectivity uplink} 
\begin{figure}
    \centering
    \includegraphics[width=\linewidth]{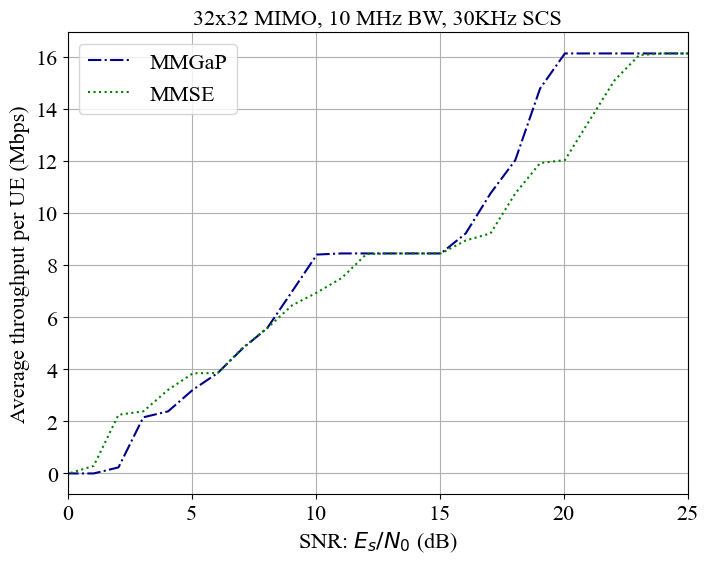}
    \caption{High connectivity uplink: $32 \times 32$ MIMO with 10 MHz bandwidth and 30 KHz SCS.}
    \label{fig:tput_high_con}
\end{figure}
Next, we evaluate the performance of \systemname{} in high-connectivity scenarios. In these scenarios, the base station is equipped with a large number of antennas and communicates with a large number of IoT UEs. Since the data rate requirements of IoT UEs are not very high, we restrict the system to use lower-order modulations only (up to 16-QAM or MCS 16) and low bandwidth. We evaluated $32\times32$ MIMO system with 10 MHz bandwidth, 30 KHz subcarrier spacing, and number of anneals $N_a = 128$. Note that in this scenario, the execution time of MMSE-SIC for a single MIMO instance exceeds the 5G timing deadline, therefore, both MMSE-SIC and \systemname{}-E cannot be used.

We see in Figure~\ref{fig:tput_high_con}, for very low SNR \systemname{}
performs worse than MMSE. This is consistent with simulation based results of prior works~\cite{di-mimo,mdimimo} which shows that at very low-SNR and BPSK modulation MMSE performs better than CIM-based appraoch. However as the SNR increases, \systemname{} can provide significant throughput improvements. Therefore practically the base station would use MMSE for low-SNR scenario and switch to \systemname{} at high SNR.






\subsection{\systemname{} Execution Time}
\label{sec:exec_time}
\begin{figure}
    \centering
    \includegraphics[width=\linewidth]{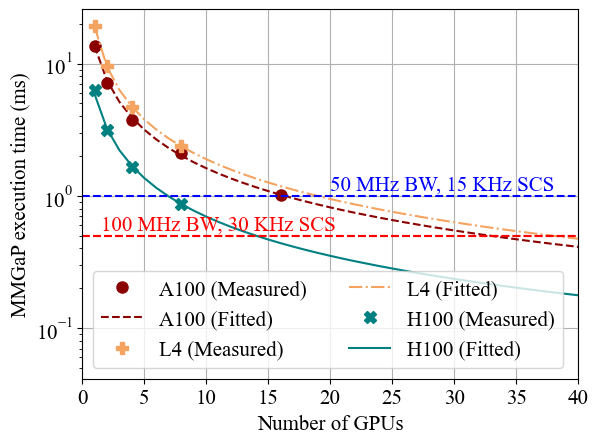}
    \caption{\systemname{} execution time for an $8 \times 8$ MIMO system with 100 MHz bandwidth and 30 KHz SCS on various NVIDIA GPUs.}
    \label{fig:gpu-runtime}
\end{figure}
In this section, we microbenchmark the execution time of our proposed solver on various NVIDIA GPUs. As described before (Figure~\ref{fig:mmgap_gpu_design}) \systemname{} divides the subcarriers  among the available GPUs (data-decomposition-based approach), all the processing is performed independently without any need of synchronization between GPUs. So, to solve $N$ MIMO problem with $K$ GPUs, we measure the time taken by $K$ OpenMP CPU threads to move the required data into the memory of $K$ GPUs respectively, run \systemname{}, and copy the result back to the CPU memory. Note that, each GPU is responsible for solving $\lfloor N/K\rfloor$ MIMO problems.

\begin{figure}
    \centering
    \includegraphics[width=\linewidth]{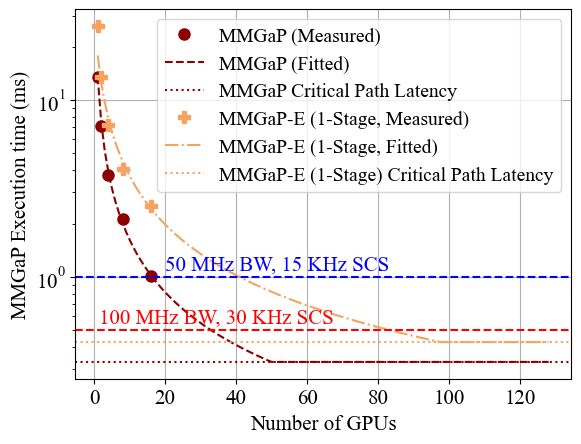}
    \caption{Comparison of \systemname{}-E and \systemname{} execution time $8 \times 8$ MIMO system with 100 MHz bandwidth and 30 KHz SCS on NVIDIA A100 GPUs. The critical path is with respect to the minimum possible work load, \textit{i.e.}, one sub-carrier and one symbol.}
    \label{fig:runtime_m_m_e}
\end{figure}

In Figure~\ref{fig:gpu-runtime}, we present the execution time of our proposed method vs. the number of GPUs. We perform measurements on three different types of GPUs: NVIDIA A100, NVIDIA L4, and NVIDIA H100 with a workload equivalent to an $8 \times 8$ MIMO system with 100 MHz bandwidth and 30 KHz subcarrier spacing. We see that \systemname{} achieves execution times in the order of milliseconds, which is required to keep up with processing demands of cellular networks. We also see that the data decomposition-based data-based employed by our system is effective and the execution times is proportional to the inverse of the number of GPUs.  With sufficient number of GPUs, \systemname{} can operate at line rate for large 5G systems, for \textit{e.g.}, 100 MHz bandwidth at 30 KHz subcarrier spacing and 50 MHz bandwidth at 15 KHz sub-carrier spacing.
Next, let us look at the total execution time of \systemname{}-E as a function of Number of NVIDIA A100 GPUs in Fig.~\ref{fig:runtime_m_m_e}. 
To measure an upper bound on the scaling of \systemname{} and \systemname{}-E, we use the concept of critical path execution time: given arbitrarily high compute resources, what is the execution time of our proposed algorithms? 
Practically, the execution time of the critical path will correspond to a scenario in which an entire A100 GPU is available to solve a single MIMO problem. Furthermore, for \systemname{}-E, the critical path measurement also assumes that both the MMSE and MMSE-SIC functions and the corresponding solver kernel runs are performed in parallel. We see that both \systemname{} and \systemname{}-E can meet the 5G processing deadlines with multiple NVIDIA A100 GPUs. We also note that \systemname{} -E (1-Stage) requires approximately twice as many GPUs as \systemname{}, which is expected as \systemname{} (1-Stage) need to run the solver kernel twice (with MMSE and MMSE-SIC guesses). 
\subsection{Discussion on Deployment Cost}
As noted in Section~\ref{sec:exec_time}, in order to meet the timing requirements of 5G, \systemname{} requires about 30 A100 GPUs for 100 MHz BW, 30 KHz SCS. Based on the pricing available on popular online retail websites (like Amazon.com), one A100 would cost around 7.5K USD. Therefore, the total price for purchasing 30 A100s would be 225K USD. For a 50 MHz and 30 KHz SCS 5G system, which requires only 15 A100s (as seen in Section~\ref{sec:exec_time}), the cost would be 112.5K USD. But how does this compare to the cost of base station?  

Although the cost of actual 5G base stations is not directly available, we can use the FCC report ``Final Catalog of Eligible Expenses And Estimated Costs''~\cite{fcc_report}, which estimates the cost of removal, replacement, and disposal of equipment from Huawei Technologies Company and ZTE Corporation. Assuming that replacement of 5G cell-site would constitute the majority component of this estimate, we can consider the estimates by FCC as a tight bound for the cost of a 5G cell-site. Based on FCC's report we see that, for 5G cell-sites, the average cost can be as high as 1.1M USD. Therefore, the addition of 30 A100 GPUs to the 5G cell site would lead to about $20\%$ increase in expenditure. Further note that we are overestimating the additional expenditure as the estimated cost of GPUs is based on retail price, but purchase of large number of GPUs by base station manufacturers would incur bulk-order pricing, which is usually heavily discounted compared to retail price. 

\section{Related Work}
\label{sec:related}

\textbf{Coherent Ising Machines and MIMO detection.} In 2013, Yamamoto \textit{et.al.} proposed an optical system based on degenerate optical parametric oscillators~(DOPO) as a CIM~\cite{wang2013coherent} and demonstrated that it could effectively solve the Ising optimization problem. Since then, several other variants of CIM~\cite{wang2013coherent,marandi2014network,mcmahon2016fully,inagaki2016coherent,dopo,oeo} have been proposed in the literature. The ability of CIMs to effectively solve NP-Hard problems attracted the attention of researchers in several domains including wireless communications, and in 2022, Singh~\textit{et. al} proposed, RI-MIMO, the first CIM-based MIMO detector~\cite{ri-mimo}. Later, Singh~\textit{et. al} developed more sophisticated CIM-based MIMO detectors (DI-MIMO and MDI-MIMO)~\cite{di-mimo,mdimimo} and showed that CIM-based MIMO detectors were capable of not only providing significant performance gains, even for systems with many users and very high modulations (like $32 \times 32$ MIMO, 256-QAM), but also provide near-theoretically optimal performance in many scenarios. However, their evaluation is limited to performance results via MATLAB simulations, without any consideration for the timing requirements of a practical system like 5G NR.

\textbf{Practical MIMO Implementations.} Several works in the literature propose practical implementations of Massive MIMO systems. Agora~\cite{agora} provides a CPU-based design for the baseband processing of a Massive MIMO base station that utilizes the AVX2 and AVX-512 SIMD instruction sets. However, Agora uses the Zero Forcing (ZF) MIMO detector, which is the simplest MIMO detector based on channel inversion. The performance of ZF deteriorates drastically as the number of antennas at the base station becomes equal to the number of users. \textit{i.e.}, the cell transition from a massive MIMO setup to a large MIMO setup. Expanding a massive MIMO system to a large MIMO allows the base station to support more users simultaneously and potentially increase the overall cell throughput~\cite{ri-mimo}. Unlike Agora, \systemname{} performs MIMO detection via a physics-inspired Ising algorithm, which can provide significantly higher throughput. Big-station~\cite{bigstation} provides a parallel implementation of Large MU-MIMO processing using commodity hardware. Argos~\cite{argos} provides a base-station implementation capable of Multi-user beam-forming. Hydra~\cite{hydra} provides a scalable, distributed computing-based design for MIMO baseband processing. LuMaMi~\cite{lumami} is another popular massive MIMO testbed that supports 100 antennas at the base station and up to 12 users (simultaneously). The design of all these testbeds, Big-station, Argos, Hydra, and LuMaMi, implements the simple ZF MIMO detector. As shown in the evaluation, \systemname{} can provide a significant improvement in throughput over the MMSE detector (which is known to be better than ZF~\cite{mmse_zf_1}). 

\textbf{Quantum annealing (QA) based MIMO detectors.} QA-based approaches include QuAMax~\cite{minsung2019}, classical-quantum hybrid MIMO detection~\cite{pSuccessQA}, and Iot-ResQ~\cite{kim2022warm}. Although these methods show promise, their gains are limited to scenarios with low-order modulations (BPSK and QPSK), and they fail to provide performance gains for higher-order modulations (16-QAM and higher). This is a major drawback, as practical systems regularly use high-order modulations. Further, the programming time (time taken to map the problem onto qubits) of quantum annealers is very high, making them infeasible for practical wireless systems. 

Paramax~\cite{minsungParallelTemp} proposes a parallel tempering-based MIMO detector that provides significant performance improvements over QA-based methods. However, even Paramax in unable to provide performance gains for high-order modulations. It has been shown that CIM-based MIMO detection vastly outperforms Paramax~\cite{mdimimo}. Furthermore, Paramax is based on a software simulation of parallel tempering algorithm, and its feasibility to meet the timing requirements of a practical 5G system remains unclear. 

Machine Learning~(ML) provides another radical alternative to conventional MIMO detectors. ML-based methodologies involve the use of Deep Neural Networks~\cite{detnet}, Gaussian Mixture Models~\cite{ML5}, Graph Neural Networks~(GNN)~\cite{neural_nr},  Recurrent Neural Networks (RNN)/LSTMs~\cite{ML1,lisa}, and Convolutional Neural Networks~(CNN)~\cite{ML0} for solving the MIMO problem. Although these methods show promise, their gains have been demonstrated primarily for small systems with few users and base station antennas and/or systems that use low-order modulations. CIM-based MIMO detection has been shown to significantly outperform machine learning-based MIMO detectors~\cite{mdimimo}.

\section{Conclusion}
\label{s:conclusion}
\systemname{} is a scalable GPU-based ``Physics-inspired'' MIMO detector that can meet the processing requirements of state-of-the-art 5G systems. It provides a radical alternative to conventional computing-based linear MIMO detectors, which have dominated practical implementations of cellular systems for more than two decades~(due to their simplicity and low complexity). We have demonstrated that \systemname{} can significantly improve the performance of linear MIMO detectors (which are popularly used by the industry), seamlessly scale to multiple GPUs, and achieve efficient data-decomposition-based parallelism. 
\systemname{} serves as a proof that ``Physics-inspired'' MIMO detection is not only practical for real-world cellular systems, but can also achieve the performance gains predicted by simulation-based studies. 
\section{Acknowledgments}
This material is based upon work supported by the National Science Foundation under grant CNS-2433915. We gratefully acknowledge a gift from Princeton NextG industrial affiliate program member Qualcomm Corporation. We also acknowledge the support of Defense Advanced Research Projects Agency (DARPA) under  grant No. FA8750-22-C-1034.

\bibliographystyle{concise2}
\let\oldbibliography\thebibliography
\renewcommand{\thebibliography}[1]{%
  \oldbibliography{#1}%
  \setlength{\parskip}{0pt}%
  \setlength{\itemsep}{0pt}%
}
\bibliography{reference}
\clearpage 
\appendix
\section{Numerical Integration Microbenchmarks}
\label{sec:num_int_micro}
\begin{figure}
    \centering
\includegraphics[width=\linewidth]{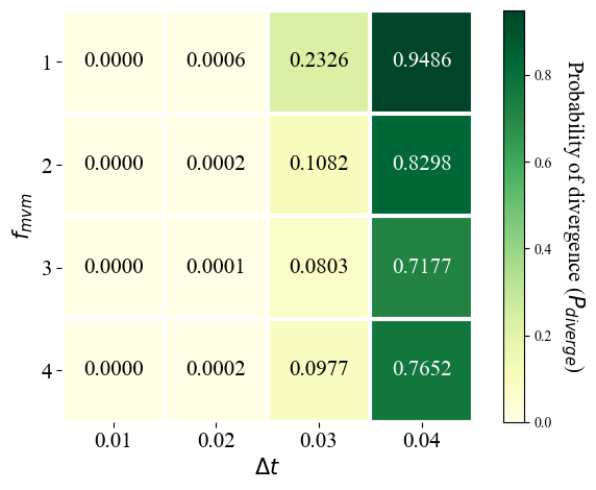}
    \caption{Heatmap illustrating the probability of divergence \textit{i.e.} the numerical integration of CIM-CAC equations fail to converge.}
    \label{fig:pdiv_heatmap}
\end{figure}

In this section, we evaluate the impact of the integration interval ($\Delta t$) and the frequency of MVM computation ($f_{mvm}$). 

Performing numerical integration with a small integration interval ($\Delta t = 0.01$) and MVM updates at each integration step~($f_{mvm} = 1$) would accurately converge to the correct steady state for the CIM-CAC equations, but it is computationally slow. However, since the output of \systemname{} only depends on the signs of the final CIM-CAC states, it does not matter whether the intermediate values of the CIM-CAC states differ from the true integration values, as long as the sign of the final states is the same. Furthermore, if $\Delta t$ is too large, then the numerical integration can fail to converge and the output is meaningless.

Therefore, to benchmark the impact of varying $\Delta t$ and $f_{mvm}$, we calculate whether the sign of the final states with modified parameters is the same as those with the high fidelity configuration ($\Delta t = 0.01$, $f_{mvm} = 1$, $N_I = 256$ integration steps). We keep the total integration length the same, \textit{i.e.}, $N_I \cdot \Delta t$ is the same for all scenarios. We define
\begin{equation}
    P_{diverge} = \frac{\text{\#instances for which output diverges}}{\text{\#total number of instances}}.
\end{equation}
\begin{equation}
    P_{error} = \frac{\text{\#state variables with error in final sign}}{\text{\#total number of state variables}},
\end{equation}

We run 10K instances with random Ising coefficients and initial values for the CIM-CAC state variables (kept constant across different parameter settings) and evaluate $P_{diverge}$ and the expected value of $P_{error}$.

\begin{figure}
    \centering
    \includegraphics[width=\linewidth]{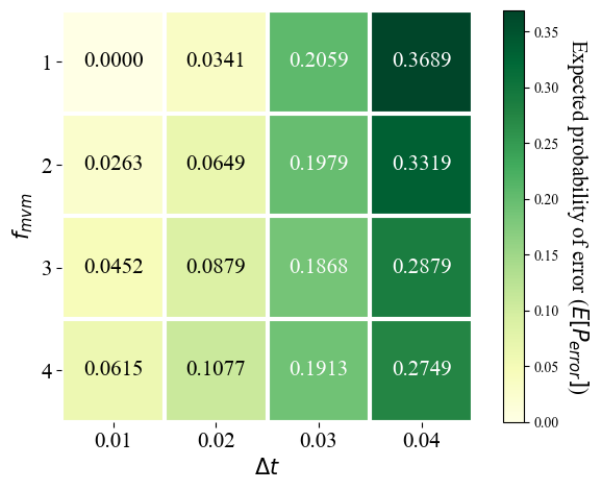}
    \caption{Heatmap illustrating the probability of error in the signs of the final state of the CIM-CAC evaluated by 
    \systemname{} vs. the ideal final state, for various values of $f_{mvm}$ and $\Delta t$}
    \label{fig:perr_heatmap}
\end{figure}

We see from Figure~\ref{fig:pdiv_heatmap} that $\Delta t$ larger than 0.02 has a very high probability of divergence and therefore is not suitable for use. Furthermore, we see from Figure~\ref{fig:perr_heatmap} that the probability of error in the signs of the final states increases with an increase in $\Delta t$ and $f_{mvm}$. Based on these empirical results, we choose $\Delta t = 0.02$ and $f_{mvm} = 2$ for \systemname{}, as this configuration provides a significant reduction in complexity while having an acceptable $P_{error}$ and $P_{diverge}$ close to zero.

\end{document}